\documentclass[12pt]{article}
\def\dm          {\ifmmode {\Delta {\rm m}^2}
                 \else    {$\Delta {\rm m}^2$}\fi}
\def\nutau       {\ifmmode {\nu_{\tau}}
                 \else    {$\nu_{\tau}$}\fi}
\def\numu        {\ifmmode {\nu_{\mu}}
                 \else    {$\nu_{\mu}$}\fi}
\def\nue         {\ifmmode {\nu_{\rm e}}
                 \else    {$\nu_{\rm e}$}\fi}
\def\elec        {\ifmmode {{\rm e}^-}
                 \else    {${\rm e}^-$}\fi}

\def\numtonue    {\ensuremath{\nu_{\mu}\rightarrow\nu_{\rm e}}}
\def\numtonut    {\ensuremath{\nu_{\mu}\rightarrow\nu_{\tau}}}

\usepackage{a4p}
\usepackage{epsf}

\begin{document}
\bibliographystyle{srt}
\begin{titlepage}
\begin{center}{\large   EUROPEAN LABORATORY FOR PARTICLE PHYSICS
}\end{center}\bigskip
\begin{flushright}
CERN-OPEN-99-010 \\
May 18, 1999
\end{flushright}
\bigskip\bigskip\bigskip\bigskip\bigskip
\begin{center}{\huge\bf
{
 \boldmath
 Study of a Long Baseline \\
 \nutau\ Appearance Neutrino \\
 Oscillation Experiment in the \\
 \vspace*{0.25cm}
 Quasi-Elastic Regime
 \unboldmath
}
}\end{center}\bigskip\bigskip
\begin{center}
  {\large Mathieu Doucet$^a$, Jaap Panman$^a$ and Piero Zucchelli$^{a,b}$}\\
          $^a$ CERN, Geneva, Switzerland \\
          $^b$ On leave of absence from INFN~-~Sezione di Ferrara
\end{center}
\bigskip\bigskip\bigskip\bigskip\bigskip
\begin{center}{\large  Abstract}\end{center}
\begin{center}\begin{minipage}{5.2truein}
We present a study for a design of a long baseline 
$\nu_{\mu} \rightarrow \nu_{\tau}$ appearance experiment to probe 
the high $\sin^22\theta$ and low $\Delta m^2$ region relevant to explain 
the atmospheric neutrino anomaly.  The experiment relies on a good 
identification of quasi-elastic interactions, which is a clean topology
that has an important contribution in the lowest $\Delta m^2$ part of 
the region probed.  The detector we studied is a fine grained liquid 
scintillator detector of 15~kilotons, optimized to detect electrons from 
$\tau\rightarrow{\rm e}\bar{\nu}_{\rm e}\nu_{\tau}$ decays, while 
rejecting backgrounds from $\pi^0$ in \numu\ interactions and electrons 
from the $\nu_{\rm e}$  beam contamination.  As a reference, the proposed 
$\nu_{\mu}$ neutrino beam from CERN to Gran Sasso was used.
\end{minipage}\end{center}
\bigskip\bigskip\bigskip\bigskip
\bigskip\bigskip
\begin{center}{\large
}\end{center}
\end{titlepage}
\section{Introduction}
The question of whether or not neutrinos have a mass has been in the mind
of physicists for some time.  If the neutrinos have non-degenerate masses 
and are mixed, neutrino oscillations could in principle be observed.
The study of this phenomenon, by which transitions between neutrino flavours 
are possible, constitutes a very precise way to determine the mass states 
of the neutrinos.

Although there have been some indications from solar neutrino experiments 
(Homestake~\cite{homestake}, GALLEX~\cite{gallex}, SAGE~\cite{sage}, 
Kamiokande~\cite{kamio_sol} and Super-Kamiokande~\cite{superk_sol}) that 
the neutrinos indeed could oscillate, the recent results from the 
Super-Kamiokande~\cite{superk_atm} atmospheric neutrino experiment gave 
the strongest indications, by showing evidence for \numu\ disappearance.  
Other atmospheric neutrino experiments have obtained results compatible 
with this hypothesis (Kamiokande~\cite{kamio_atm}, MACRO~\cite{macro} 
and SOUDAN2~\cite{soudan2}).  Furthermore, the LSND~\cite{lsnd} 
short-baseline appearance experiment found indications for \numu\ 
oscillating to \nue.  

Now that the first indications for neutrino oscillations have been found,
the confirmation of the oscillation signals and the determination of the
oscillation parameters represent important objectives.  This year already, 
the K2K~\cite{k2k} long baseline disappearance experiment will shed some 
light on the situation. At the end of the data taking, it will be able to 
probe the region above $\dm \sim 2\times10^{-3}$~eV$^2$.  The 
MINOS~\cite{minos} experiment at Fermilab, which will start taking data 
early next century, also aims to probe the atmospheric region.

Given the CHOOZ~\cite{chooz} result, which excludes the interpretation of 
the atmospheric neutrino data as $\numtonue$ oscillations above 
$\dm \sim 10^{-3}$~eV$^2$, a plausible interpretation of these data is 
$\numu\rightarrow\nutau$ oscillations.  Proposals for several 
experiments to detect \nutau\ appearance using the proposed 
NGS~\cite{cern9802} neutrino beam from CERN to Gran~Sasso are being prepared. 

Prompted by the recent efforts, both at Fermilab and at CERN, to develop 
long baseline projects with the NuMi~\cite{minos} and NGS beams, the purpose 
of the present paper is to study the possibilities for a simple \nutau\ 
appearance experiment to probe the \dm\ region of the atmospheric neutrino 
experiments.  We argue that a totally active scintillator detector in a 
long baseline beam would provide an efficient way to verify the 
oscillation claim from the Super-Kamiokande data.  

The present study is based on the parameters of the proposed NGS beam.  
This beam has a mean energy of about 18~GeV and a baseline of about 735~km, 
corresponding to the distance between CERN and the Gran Sasso laboratory. 
In order to understand  the role of the baseline, we also study the 
performances of the same neutrino detector when positioned 3000~km away 
from the same neutrino source.

\begin{figure}[ht]
      \begin{center}\mbox{\epsfxsize=8cm\epsffile{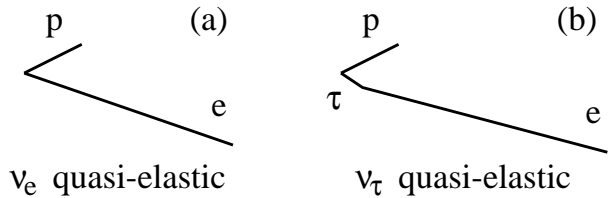}}
      \end{center}
      \caption{Schematic comparison of $\nue {\rm n} \rightarrow {\rm ep}$
               and $\nutau {\rm n} \rightarrow \tau {\rm p}$ with
               $\tau \rightarrow {\rm e} \bar{\nu}_{\rm e} \nutau$ 
               quasi-elastic processes.}
      \label{fig-topo}
\end{figure}

In these conditions, low \dm\ values (below $5\times10^{-3}$~eV$^2$) are 
such that the oscillated $\nutau$ energy spectrum peaks at low energy 
(below 10~GeV).  At these energies, the quasi-elastic $\nutau$ interactions
have a significant contribution to the total event rate.  An experiment 
aiming to reach very low values of \dm\ could thus be optimized for the 
detection of quasi-elastic processes, since they represent a very clean 
signal compared to the deep-inelastic processes.  An efficient way of 
studying those interactions is to focus on the events where the produced  
$\tau$ decays to an electron~\footnote{The branching ratio of the $\tau$ 
decay to electron is about 18\%~\cite{pdg}.} ( $\nutau {\rm n} \rightarrow 
\tau {\rm p}$, with  $\tau \rightarrow {\rm e} \bar{\nu}_{\rm e} \nutau$ ).
The signature of these interactions is similar to the \nue\ quasi-elastic 
( $\nue {\rm n} \rightarrow {\rm ep}$ ) events that I216~\cite{i216} aims 
to study to search for \numtonue\ appearance.  Figure~\ref{fig-topo} shows 
the two topologies involved for \numtonue~(a) and \numtonut~(b) appearance 
experiments.  The I216 experiment would be looking for \nue\ appearance 
in a \numu\ beam by comparing the ratio of \numu\ and \nue\ quasi-elastic
interactions in detector modules at two locations (130~m and 885~m from
the target).  The current design of the modules consists in fine grained 
fully active scintillator calorimeters amounting to an overall mass of 
500~tons.  This experiment would use the CERN-PS to get a neutrino beam of 
energy of about 1.5~GeV.  In the present study, we use one of the detector 
designs of I216 (to be described later) and apply it to the detection of 
\nutau\ interactions.

Although addressing the question of neutrino oscillations in the 
atmospheric region is of great importance, the verification of the 
oscillation claim of LSND is no less important.  The KARMEN~\cite{karmen}
experiment is trying to clarify the issue, and other projects have 
been proposed (MiniBoone~\cite{miniboone} and I216 itself).  According 
to this study, we conclude that the use of a single detector technology 
could allow to search for neutrino oscillations both in the LSND region 
and the atmospheric region by doing a short baseline \numtonue\ experiment 
as a first step to a long baseline \nutau\ appearance experiment.

This paper is organized in the following way.  A description of the
detector is given in section~\ref{sect-det}.  The event selection and the 
backgrounds are discussed in section~\ref{sect-bck}, which is followed by 
section~\ref{sect-osc} on the oscillation sensitivity. 

\section{Description of the detector}
\label{sect-det}
In the studies related to the I216 Letter of Intent~\cite{i216}, it was 
realized that a fully active liquid scintillator detector with a 
granularity of the order of a few centimetres fulfils the general 
requirements for electron identification and topological reconstruction 
necessary for a \nue\ appearance search.  We want to apply the same
technique to the detection of 
$\nutau {\rm n} \rightarrow \tau {\rm p}$ interactions with subsequent 
decay $\tau \rightarrow {\rm e} \bar{\nu}_{\rm e} \nutau$, in a detector 
about thirty times larger (15~ktons).  It should be noted that liquid 
scintillator detectors have already been successfully used in a wide 
range of experiments (for example: CHOOZ~\cite{chooz} and MACRO~\cite{macro}),
and foreseen for future large scale applications (for example:
KamLAND~\cite{kamland} and Borexino~\cite{borexino}).

\subsection{Detector design}
\begin{figure}[ht]
      \begin{center}\mbox{\epsfxsize=8cm\epsffile{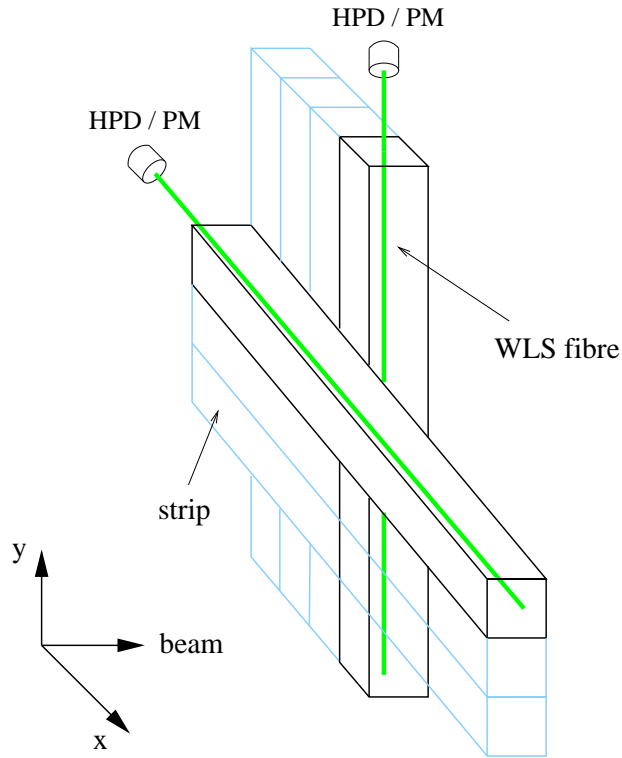}}
      \end{center}
      \caption{Schematic view of the detector planes }
      \label{fig-det}
\end{figure}

Because of the large scale, a simple and modular structure should be 
adopted (see figure~\ref{fig-det}).  The active target consists of 
an oil based scintillating mixture contained in a large vessel.  The 
granularity is provided by optical separators immersed in the liquid.  
These separators are structured in planes of parallel strips.  The 
planes are perpendicular to the beam direction.  The strips of a given 
plane are orthogonal to those of the adjacent planes in order to be 
able to do bi-dimensional reconstruction.  The light collection is 
accomplished by wavelength-shifting (WLS) fibres placed inside each 
strip.  The light from the fibres is readout by multi-pixel photon detectors.  
The general detector characteristics are reported in table~\ref{tab-det} 
and discussed in the following sub-sections.

\begin{table}[h]
  \begin{center}
    \caption{General features of the detector}
    \vspace{0.25cm}
    \begin{tabular}{|l|c|}
      \hline
 Mass                      &   15~ktons \\ \hline
 Modules                   &   4        \\ \hline
 Module width              &   10~m     \\ \hline
 Module length             &   42~m     \\ \hline
 Sampling                  &   4~cm     \\ \hline
 Granularity               &   4~cm     \\ \hline
 Planes per module         &   1042     \\ \hline
 Strips per module         & 260417     \\ \hline
 Photodetectors per module & 2035       \\ \hline
    \end{tabular}
    \label{tab-det}
  \end{center}
\end{table}

\subsubsection{The active medium}
Mineral oil scintillator has been studied in great detail, and excellent
performances in stability, light yield and attenuation lengths have already 
been reached in the past.  Contrary to surface readout detectors (for 
instance KamLAND and Borexino), where the attenuation length of the
liquid plays an important role because the light must travel a large
distance, the light emitted by the scintillator in the case of a WLS
fibre readout is captured locally inside the fibre.  The attenuation 
length of the liquid can then be shorter, which gives more freedom
in the choice of the mixture.

A very recent development of oil based liquid scintillator consists in 
using new solvents like PXE (phenyl-o-xylylethane)~\cite{koch} and 
LAB (linear alkylbenzenes)~\cite{eljen}.  Mixtures using these solvents
can have very good scintillation properties (for example, 87\% of
anthracene light output can be reached with the Bicron scintillator 
BC599-16~\cite{majewski}).
Their physical properties are well suited for a very large 
scintillating WLS fibre detector. They are non-toxic, non-flammable 
(flash point above 145$^o$) and good insulators.  Both solvents 
mentioned here are produced in large quantities for industrial 
applications.  These physical properties are crucial if the 
photodetectors are to be placed inside the scintillating mixture without 
any specific precaution concerning high voltages and power dissipation. 

\subsubsection{Granularity}
In this study, we fix the longitudinal sampling frequency to 4~cm 
(corresponding to 1/13 radiation length) with a transverse granularity
of 4~cm, to allow for topological reconstruction of the events. 
With such a granularity, the detector could be placed above ground.
According to simulation and laboratory tests~\cite{minospvc}, the proper
reflectivity of the walls can be obtained in various ways and with 
different materials (for example: aluminium painted with $TiO_2$, or 
extruded polypropylene).

\subsubsection{Wavelength shifting fibres}
The WLS fibre readout technology allows to collect the scintillating 
light from a large volume (the volume of a scintillator strip) with 
good efficiency onto a small surface (the fibre diameter at one end). 
This is crucial in our large scale application, because a significant 
fraction of the overall cost is given by the  photodetectors. It is 
then important to maximize the amount of active mass which is read by 
a given photocathode surface.  For this reason, we want to have strips 
as long as possible while having enough photoelectrons to measure the 
ionization loss.  We assume a strip length of 10~meters and a 
single-sided readout with a mirror on the opposite side of the fibre.  
From studies done for MINOS~\cite{minos}, we expect that a measured 
signal of an average of 10~photoelectrons~\footnote{We assume a typical 
quantum efficiency for the photodetector (see next paragraph).} for a 
minimum ionizing particle passing through the far end of a strip could 
be achieved with current technology. 

\subsubsection{Photodetectors}
Multi-anode phototubes and hybrid-photodiodes (HPD) are commercially
available.  They allow to read a large number of channels, up to a few 
hundreds, on a single device.  In the case of HPD's, the cost is mainly 
proportional to the surface of the photocathode rather than to the number 
of channels.  Typical quantum efficiencies at a wavelength of 520~nm 
(the peak value of the emission spectrum of the WLS fibre) is in the 
region of 13\% to 20\%.  We have checked that these commercial 
photodetectors can be immersed in the mentioned liquid mixtures.  This 
aspect is crucial to minimize the length of the WLS fibres outside the 
strip and the consequent light losses.

\subsubsection{Infrastructure, civil engineering and cost}

The feasibility of a 15~kiloton~detector also depends on civil engineering 
issues like safety, assembling, environmental impact and design costs. 
Industrial oil containers~\footnote{Among these, the use of an oil
tanker is a possibility.} have capacities largely exceeding our requests
and would be suited for our application.  These could then be recycled
for other purposes after the experiment is finished.  The overall 
detector can be split into four modules of about 10$\times$10$\times$40~m$^3$.
An estimate of the costs of the main items of such a detector is given in 
table~\ref{tab-cost}, based on informal contacts with manufacturers.
The containers and the infrastructure are not included.

\begin{table}[h]
  \begin{center}
    \caption{Estimate of the cost of the detector}
    \vspace{0.25cm}
    \begin{tabular}{|l|c|c|c|}
      \hline
    Item             & Unit cost [CHF] & Amount   & Price [MCHF]\\ \hline\hline
 Photodetectors      & 14.0/channel    & 1041667  & 14.6 \\ \hline
 WLS fibres          & 1.0/meter       & 10417~km & 10.4 \\ \hline
 Liquid Scintillator & 1.0/kg          & 15~ktons & 15.0 \\ \hline
 Electronics         & 10.0/channel    & 1041667  & 10.4 \\ \hline\hline
 Total               & \multicolumn{2}{c}{ }      & 50.4 \\ \hline    
    \end{tabular}
    \label{tab-cost}
   \end{center}
\end{table}

\section{Event selection and background rejection}
\label{sect-bck}
The flux components of the NGS neutrino beam are reported in 
table~\ref{tab-beam}.  The expected event rates for pure quasi-elastic,
resonances and deep inelastic processes~\footnote{We define the 
deep-inelastic processes as those which occur above $W^2=2$~GeV$^2$.}
were computed using the latest version of the NGS neutrino beam~\cite{vass}  
(they are given in table~\ref{tab-rates}).  The number of \nutau\ 
interactions for several values of~\dm\ are shown.  The rates are given 
per kiloton assuming four years of NGS beam at $4 \times 10^{19}$ protons 
on target per year.

\begin{table}[h]
  \begin{center}
    \caption{Neutrino beam components at the Gran Sasso location.  The numbers
             indicated with a $\star$ are those of the contamination
             quoted in Reference~\cite{cern9802}.  The other values are those
             of the latest version of the NGS beam~\cite{vass}}
    \vspace{0.25cm}
    \begin{tabular}{|l|c|c|}
      \hline
 Source                      & total flux [$\nu/({\rm pot\;m}^2)$] 
                                   & relative importance    \\ \hline\hline
 $\nu_{\mu}$                 & $7.2\times10^{-9}$  & 100\%            \\ \hline
 $\bar{\nu}_{\mu}$           & $1.6\times10^{-10}$ & 2.2\%$^{\star}$  \\ \hline
 $\nu_{\rm e}$               & $4.4\times10^{-11}$ & 0.6\%            \\ \hline
 $\bar{\nu}_{\rm e}$         & $5.8\times10^{-12}$ & 0.08\%$^{\star}$ \\ \hline
    \end{tabular}
    \label{tab-beam}
   \end{center}
\end{table}

From table~\ref{tab-rates}, it can be seen that the quasi-elastic 
processes for oscillated \nutau\ are a significant fraction of the 
total number of \nutau\ interactions.  Furthermore, the resonant 
interactions, which also make an important fraction of the total 
number of events, can also have a topology similar to the quasi-elastic 
processes.  Therefore, we focus our analysis on the clean quasi-elastic 
topology where the $\tau$ decays to an electron. We define this 
topology as an electromagnetic shower and at most one additional 
charged track in the final state.  We shall see how it is possible to 
have a good efficiency by a simple selection of these events while 
rejecting an important fraction of the background coming from the 
\nue\ contamination of the beam as well as the other sources of background.

The selection of events could be refined by studying other event
properties such as missing transverse momentum.  Crucial experimental
data giving information about the properties of the background, including
cross-section measurements, could in principle come from another 
experiment like I216, where \nutau\ oscillations are excluded at large
mixing angle.

It should be mentioned that other decay channels of the $\tau$ can be 
studied in addition to the electronic channel.  For instance, the 
$\tau \rightarrow \pi^-\nutau$ decay in quasi-elastic interactions was 
studied in CHARM~II~\cite{charm2} for a \numtonut\ oscillation search
at high sensitivity.  This channel could provide a signal independent 
of the \nue\ contamination of the beam.

\begin{table}[h]
  \begin{center}
    \caption{Expected number of events per kiloton at 735~km for 
             different processes 
             for four years of NGS beam at $4 \times 10^{19}$ pots/year.  
             The numbers in parentheses are the numbers of events with a
             $\tau$ decaying to an electron for the quasi-elastic (QE)
             and resonant (RES) processes.  All processes are 
             charged-current interactions except where mentioned (NC)}
    \vspace{0.25cm}
    \begin{tabular}{|l|c|c|}
      \hline
 Source                 & Events/kton     \\ \hline\hline
\nutau\ QE [$\dm =5.0\times10^{-3}$ eV$^2$]  & 25.2 (4.49)  \\ \hline
\nutau\ QE [$\dm =2.5\times10^{-3}$ eV$^2$]  & 6.79 (1.21)  \\ \hline
\nutau\ QE [$\dm =1.5\times10^{-3}$ eV$^2$]  & 4.39 (0.78)  \\ \hline
\nutau\ QE [$\dm =1.0\times10^{-3}$ eV$^2$]  & 1.11 (0.20)  \\ \hline\hline

\nutau\ RES [$\dm =5.0\times10^{-3}$ eV$^2$] & 53.2 (9.47)  \\ \hline
\nutau\ RES [$\dm =2.5\times10^{-3}$ eV$^2$] & 14.7 (2.61)  \\ \hline
\nutau\ RES [$\dm =1.5\times10^{-3}$ eV$^2$] & 9.49 (1.69)  \\ \hline
\nutau\ RES [$\dm =1.0\times10^{-3}$ eV$^2$] & 2.41 (0.43)  \\ \hline\hline

\nutau\ DIS [$\dm =5.0\times10^{-3}$ eV$^2$] & 156          \\ \hline
\nutau\ DIS [$\dm =2.5\times10^{-3}$ eV$^2$] & 40.0         \\ \hline
\nutau\ DIS [$\dm =1.5\times10^{-3}$ eV$^2$] & 14.5         \\ \hline
\nutau\ DIS [$\dm =1.0\times10^{-3}$ eV$^2$] & 6.45         \\ \hline\hline
                                                           
 \numu\ QE+RES                               & 1060         \\ \hline
 \numu\ RES (NC)                             & 190          \\ \hline
 \nue\ QE+RES                                & 6.4          \\ \hline
 \numu\ DIS                                  & 8320         \\ \hline
 \nue\ DIS                                   & 74           \\ \hline
    \end{tabular}
    \label{tab-rates}
   \end{center}
\end{table}

\begin{figure}[ht]
      \begin{center}\mbox{\epsfxsize=14cm\epsffile{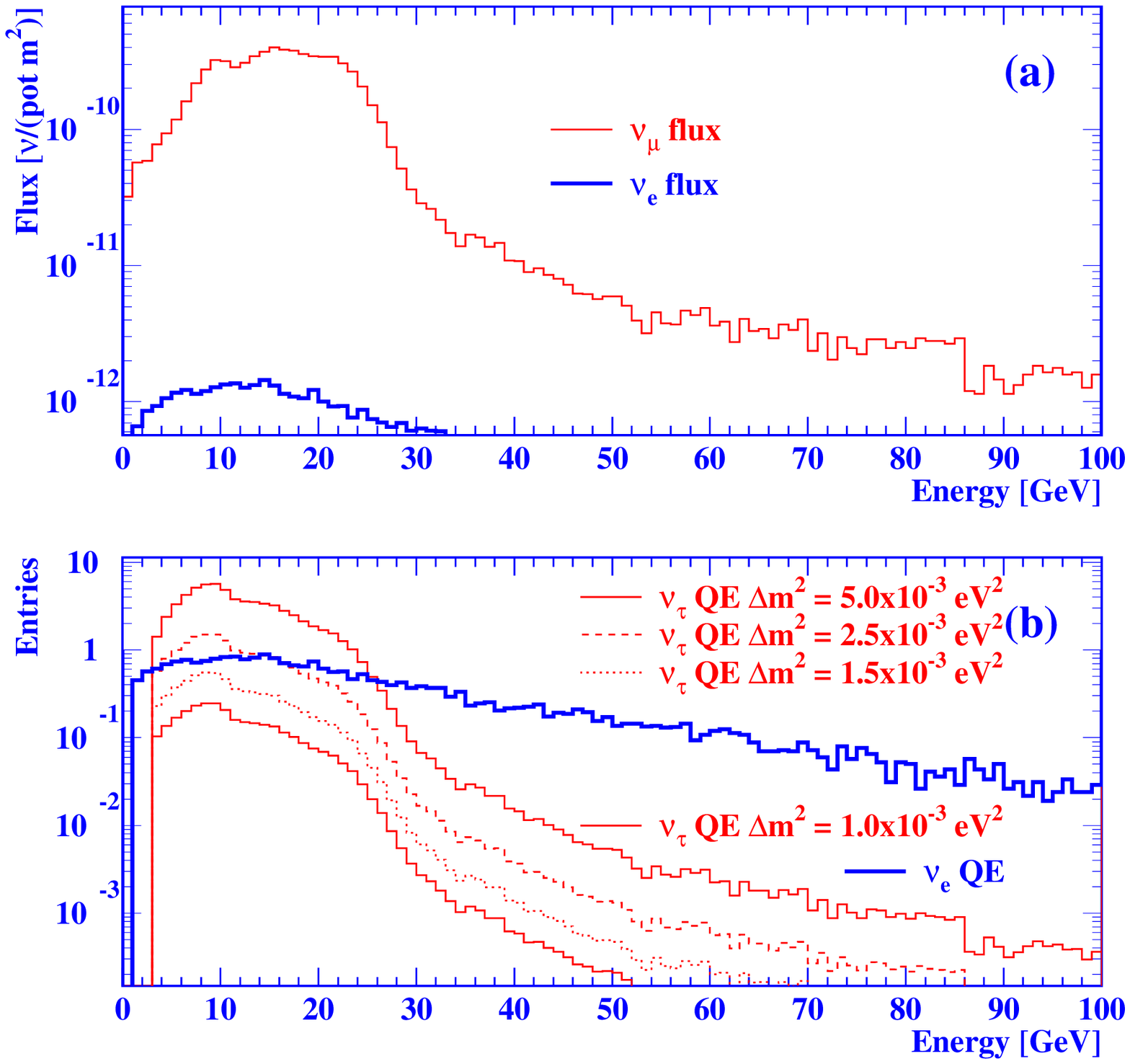}}
      \end{center}
      \caption{Neutrino flux of the NGS beam at the far location~(a).
               Neutrino energy distribution~(b) of \nue\ quasi-elastic events
               and \nutau\ quasi-elastic events where the $\tau$ decays to 
               an electron, assuming different values of \dm. 
               The normalization of figure~b is the same as in 
               table~\ref{tab-rates}}
      \label{fig-flux}
\end{figure}

\begin{figure}[ht]
      \begin{center}\mbox{\epsfxsize=14cm\epsffile{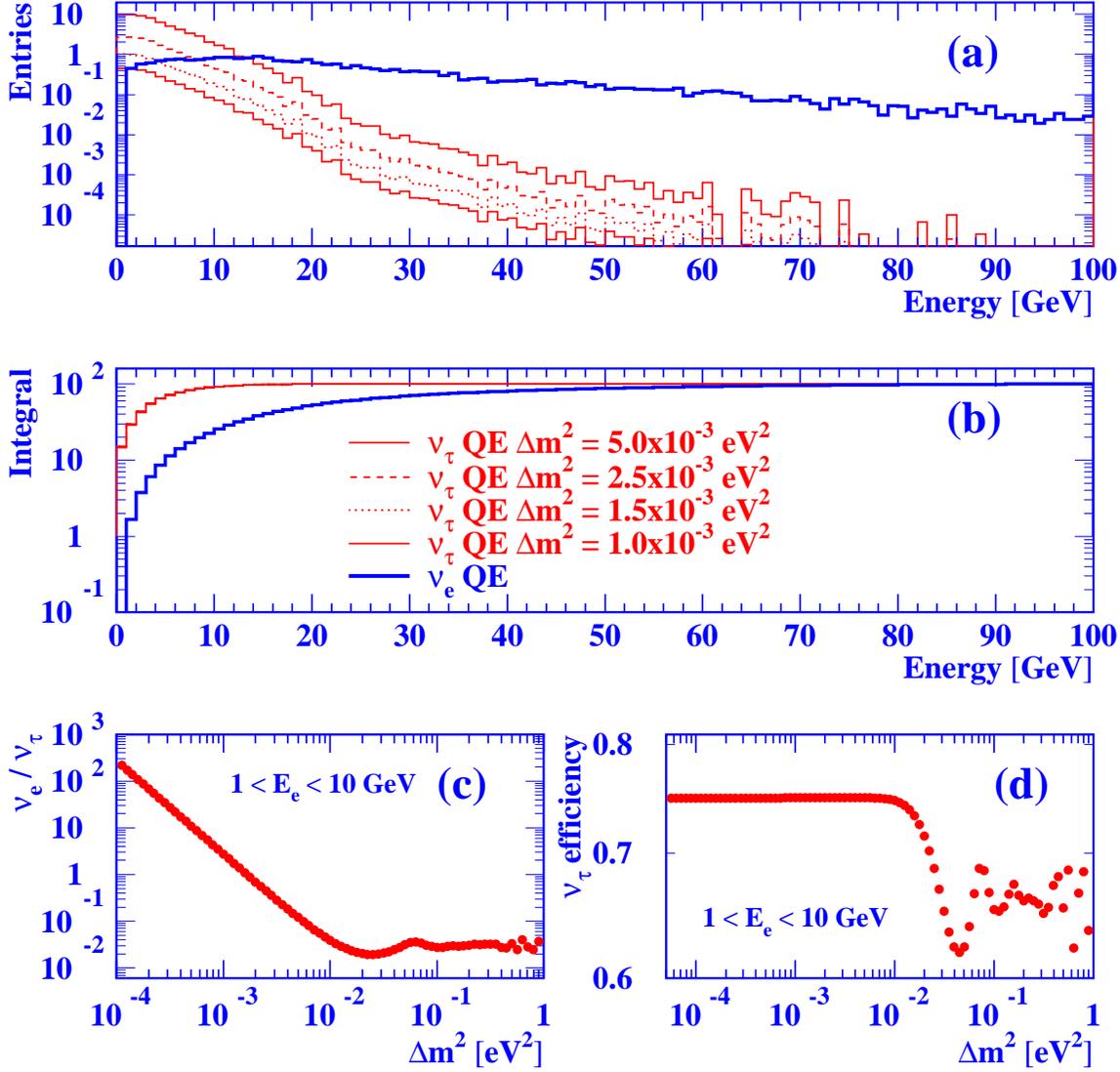}}
      \end{center}
      \caption{Energy distribution of the electron in \nue\ 
               quasi-elastic events
               and in \nutau\ quasi-elastic events where the $\tau$ decays to 
               an electron (a), integral of the distributions of (a) as a 
               function of the energy (b), ratio of background to signal for
               a cut at 10~GeV as a function of \dm\ (c), efficiency of the
               10~GeV cut selection as a function of \dm }
      \label{fig-elec_energy}
\end{figure}

\begin{figure}[ht]
      \begin{center}\mbox{\epsfxsize=14cm\epsffile{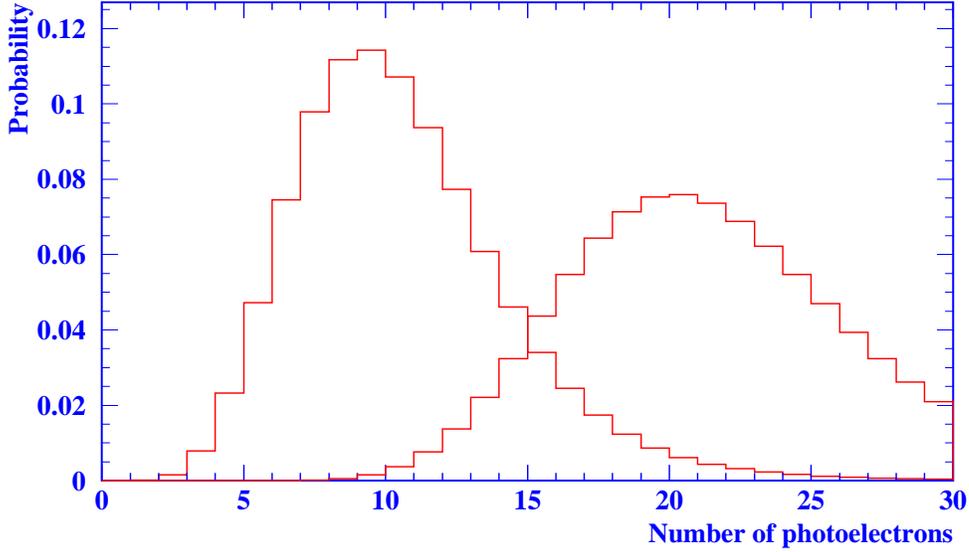}}
      \end{center}
      \caption{Probability distribution of the number of photoelectrons
               assuming an average of 10~photoelectrons for a minimum 
               ionizing particle crossing a single strip.  The peak
               at lower values is for a single particle while the 
               peak at higher values is for two minimum ionizing particles}
      \label{fig-mip}
\end{figure}

Several processes can produce events that mimic the quasi-elastic
topology of the signal.  The following contaminations were studied:

\begin{itemize}
\item{\underline{\bf \boldmath \nue\ \unboldmath contamination}\\
The main source of background for an appearance experiment selecting
quasi-elastic \nutau\ interactions where the $\tau$ decays to an 
electron is the \nue\ contamination of the beam.  As can be seen in 
table~\ref{tab-beam}, this contamination is of the order of 0.6\% of 
the \numu\ component.  Figure~\ref{fig-flux}(a) shows the spectrum 
of \numu\ and \nue\ of the NGS beam.

Figure~\ref{fig-flux}(b) shows the neutrino energy distributions of \nue\ 
quasi-elastic events and of \nutau\ quasi-elastic events where the 
$\tau$ decays to an electron, assuming different values of \dm.  
It can be seen that the neutrino energy distribution of the background 
from the \nue\ contamination of the beam is harder than the one
of the oscillated \nutau\ quasi-elastic events.  

One method to reject this background is to cut on the energy of the
identified electron.  Since the oscillation probability favors the low energy 
range of the \numu\ spectrum, and since the electron of the signal comes 
from the decay of a $\tau$, the electrons of the signal will have a much 
lower energy than the electrons of pure quasi-elastic \nue\ processes.  
Figure~\ref{fig-elec_energy}(a) shows the electron energy distributions 
from both quasi-elastic \nue\ interactions and $\tau$ decays from \nutau\ 
quasi-elastic interactions.  Figure~\ref{fig-elec_energy}(b) shows the
normalized integral of these distributions between zero and a given value 
of the electron energy (x-axis).  A cut at 1~GeV, which will be described 
later in the text, has been applied before integrating. Since the energy 
distributions of the electron are similar for all the \dm\ values 
considered, the four curves of the signal are overlapping and cannot be
distinguished.  Another important feature of this figure is that the 
integral of the signal rises very rapidly at low energy whereas the 
background rises slowly.  A cut on the energy of the electron 
can thus reduce the background efficiently while keeping most of the 
signal.  For instance, for a cut at 10~GeV, 75\% of the signal is kept 
while 70\% of the background is rejected.  Figure~\ref{fig-elec_energy}(c) 
shows the ratio of the number of background events and the number of 
signal events as a function of \dm, for a cut at 10~GeV.  
Figure~\ref{fig-elec_energy}(d) shows the signal selection efficiency 
for a 10~GeV cut as a function of \dm.  The final choice of the energy 
cut should be done by maximizing the sensitivity of the experiment, which 
will critically depend on the choice of other event selection criteria, 
baseline and detector mass.  We have checked {\sl a posteriori} that 
the sensitivity does not change significantly if the maximum allowed 
lepton energy varies in the interval between 4 and 12~GeV (with a 735~km 
baseline and a 15~kiloton detector).  Therefore, in the present study, we 
fix the upper limit of the lepton energy to be 10~GeV independently of 
the baseline.  We conclude from figure~\ref{fig-elec_energy} that a good 
rejection of quasi-elastic \nue\ events can be achieved by a simple 
energy cut.  Additional criteria to reject the \nue\ contamination could 
be based on the kinematics of the events (for instance, the angle of 
the lepton with respect to the beam), but have not been used in this analysis.
}

\newpage
\item{\underline{\bf \boldmath $\pi^0$ \unboldmath in NC processes}\\
The conversion of the photons coming from a $\pi^0$ decay in a 
neutral-current process can fake the electron signature.  To reject this
background, we apply a combination of topological cuts on the signal 
recorded in the scintillator.  Assuming that the vertex position is 
known, a $\pi^0$ can mimic an electron in the following conditions:\
\begin{itemize}
\item{at least one photon converts in a strip adjacent to the vertex position, 
      while the two electromagnetic showers of the photons overlap.}
\item{one photon converts in a strip adjacent to the vertex position, while
      the other is missed (for example in an asymmetric decay of the 
      $\pi^0$). }
\end{itemize}
In the cases stated above, the pulse height of the signal recorded
in the scintillator can be used to discriminate between the passage
of a single electron (which will behave as a minimum ionizing particle
in the first tenths of a radiation length) and two electrons coming
from a photon conversion.  With a statistics of 10 photoelectrons per 
minimum ionizing particle passing through a strip, it is possible to 
reject 92\% of the two-electron background using the information of 
the first strip only (see figure~\ref{fig-mip}).  The selection 
efficiency of the signal is then about 86\%.}
\end{itemize}

Given the previous considerations, the selection of \nutau\ events 
with a quasi-elastic topology where the $\tau$ decays to an electron 
is the following:

\begin{enumerate}
\item{The event must have a pure quasi-elastic topology, with 
      a visible vertex identified by an electromagnetic shower
      and at most one charged track (assumed to be the proton).
      In the case where the proton track is not seen, we require
      a large pulse height corresponding to a heavily ionizing
      stopping particle.  This peak will then be identified as the vertex.
      To ensure a well defined topology, 
      we require that the proton track should not interact.}

\item{The electromagnetic shower should be connected to the vertex.}

\item{The visible energy of the electromagnetic shower should be 
      between 1~GeV and 10~GeV to ensure that it comes from a $\tau$ 
      decay, and not from a \nue\ quasi-elastic event.  The lower cut 
      on the energy also cuts $\pi^0$ events, which peak at low energy.}

\item{The energy deposit of the electromagnetic shower in the 
      first plane should be compatible with a single minimum ionizing
      particle.}
\end{enumerate}

\begin{figure}[ht]
      \begin{center}\mbox{\epsfxsize=14cm\epsffile{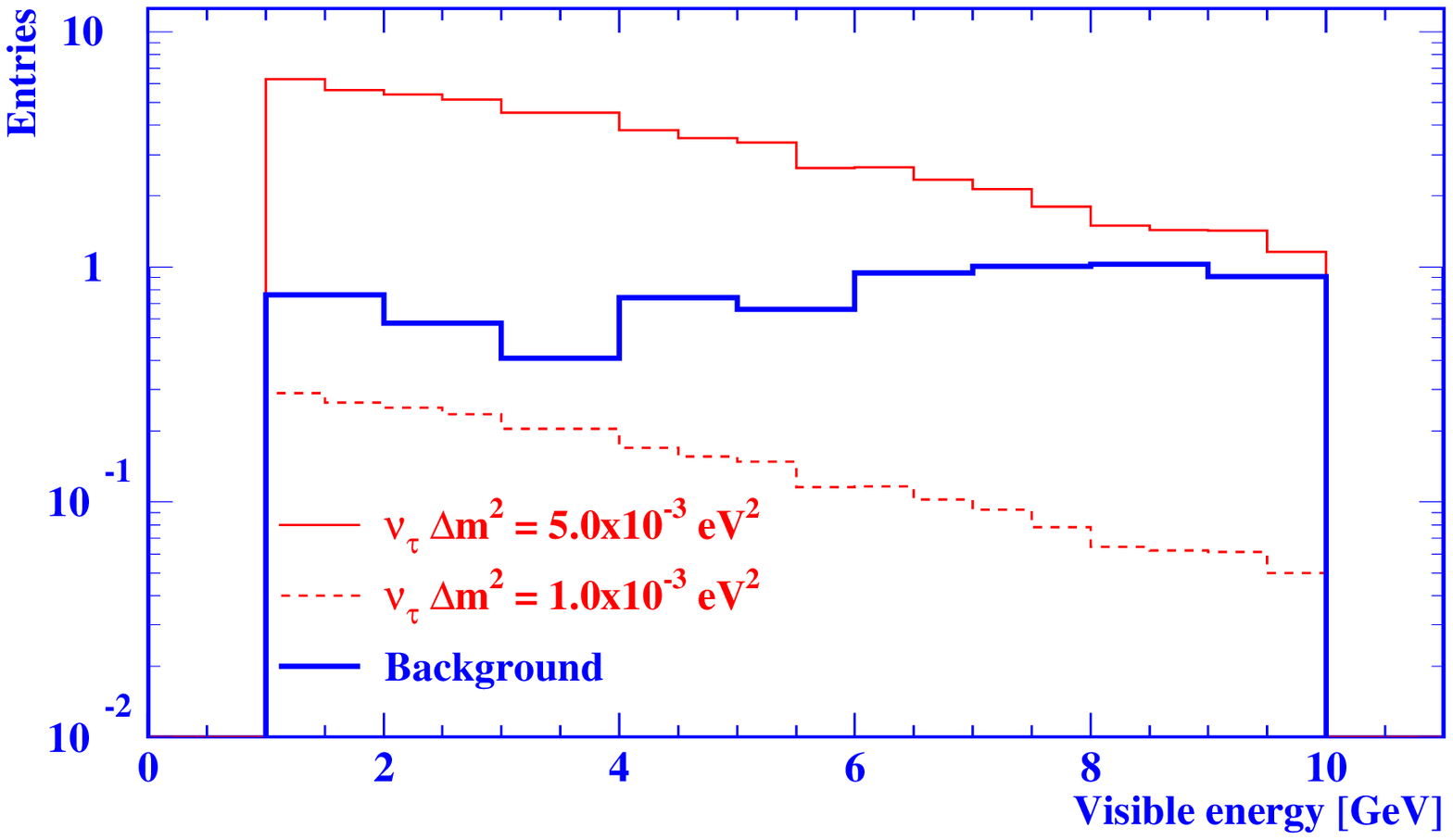}}
      \end{center}
      \caption{Visible energy for the signal and the background at 735~km}
      \label{fig-e_735}
\end{figure}

\begin{figure}[ht]
      \begin{center}\mbox{\epsfxsize=14cm\epsffile{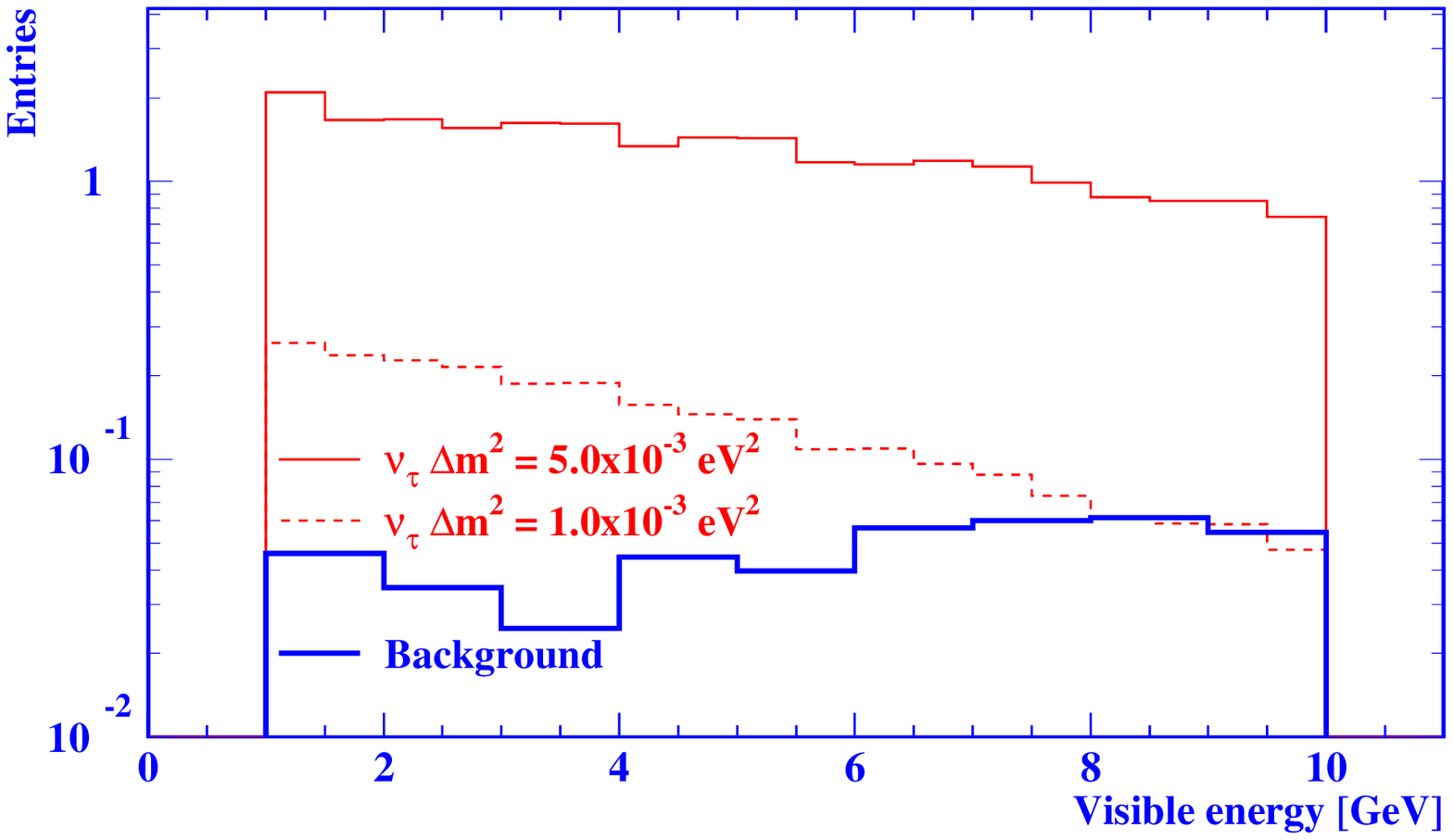}}
      \end{center}
      \caption{Visible energy for the signal and the background at 3000~km}
      \label{fig-e_3000}
\end{figure}

\begin{figure}[ht]
      \begin{center}\mbox{\epsfxsize=14cm\epsffile{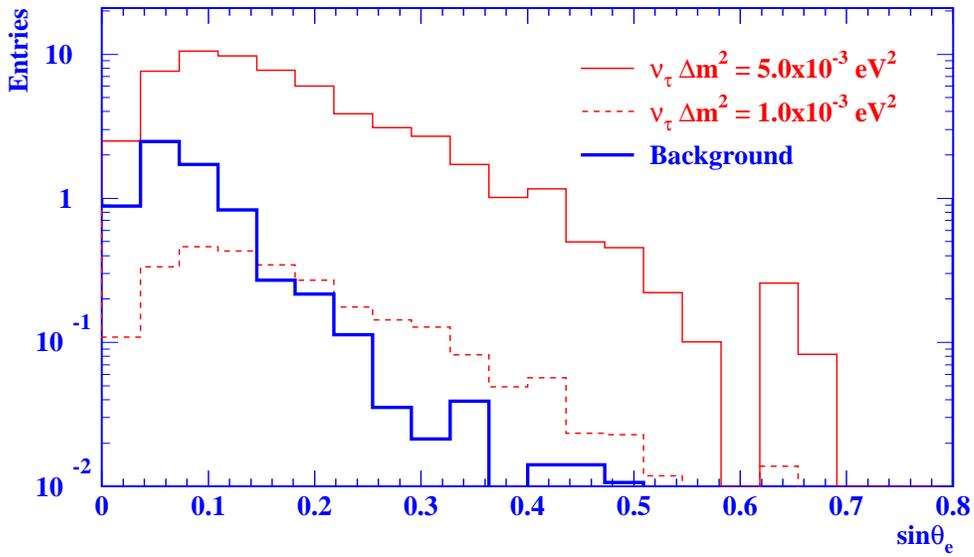}}
      \end{center}
      \caption{Angular distribution of the lepton with respect to the beam
               direction for the signal and the background at 735~km}
      \label{fig-th_735}
\end{figure}

The selection efficiencies for signal and background were evaluated
using a quasi-elastic and resonant interaction generator~\cite{resque}
and a Monte Carlo parametrized to simulate the detector response and
reconstruction.  Table~\ref{tab-rates_with_cuts} shows the different rates 
of events as computed using the cuts mentioned above. The energy 
distribution of the selected events is shown in figures~\ref{fig-e_735} 
and~\ref{fig-e_3000} for the two baselines under study. Additional
criteria could still be explored, as mentioned earlier.  As an example, 
figure~\ref{fig-th_735} shows the angular distribution of the lepton
with respect to the beam direction for the signal and the background.
\begin{table}[h]
  \begin{center}
    \caption{Expected number of events for different processes at 735~km for 
             four years of NGS beam running at $4 \times 10^{19}$ pots/year 
             assuming the efficiencies discussed in the text} 
    \vspace{0.25cm}
    \begin{tabular}{|l|c|c|}
      \hline
 Source                                    & Events/kton & efficiency \\ \hline\hline
\nutau\ QE+RES [$\dm =5.0\times10^{-3}$ eV$^2$] & 3.91   & 28\% \\ \hline
\nutau\ QE+RES [$\dm =2.5\times10^{-3}$ eV$^2$] & 1.07   & 28\% \\ \hline
\nutau\ QE+RES [$\dm =1.5\times10^{-3}$ eV$^2$] & 0.69   & 28\% \\ \hline
\nutau\ QE+RES [$\dm =1.0\times10^{-3}$ eV$^2$] & 0.18   & 28\% \\ \hline\hline
  \nue\ QE+RES                                  & 0.44   & 6.9\%   \\ \hline
 \numu\ RES (NC)                                & 0.042  & 0.022\% \\ \hline
    \end{tabular}
    \label{tab-rates_with_cuts}
   \end{center}
\end{table}

\section{Sensitivity to neutrino oscillations}
\label{sect-osc}
\begin{figure}[ht]
      \begin{center}\mbox{\epsfxsize=14cm\epsffile{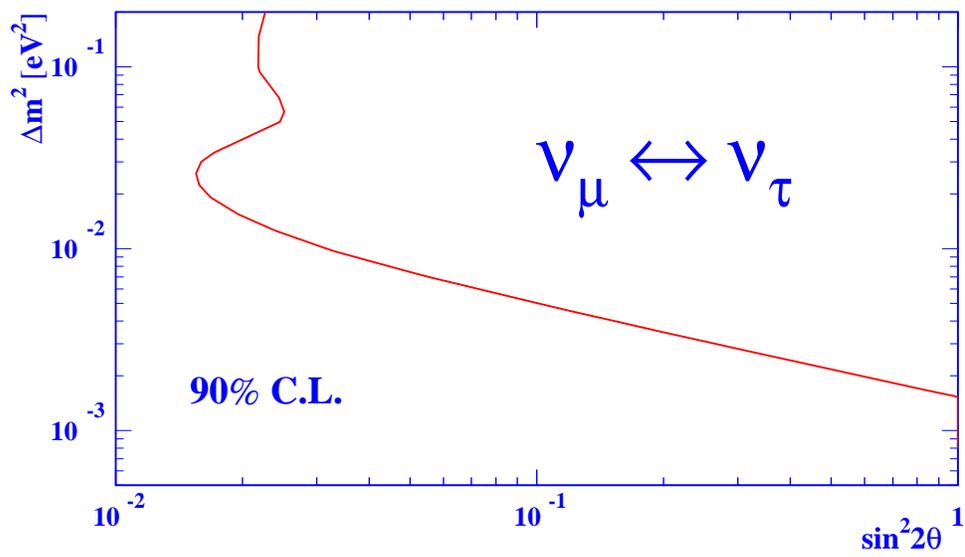}}
      \end{center}
      \caption{Sensitivity plot at 735~km}
      \label{fig-exclu1}
\end{figure}

The sensitivity is the ``average upper limit that would be
obtained by an ensemble of experiments with the expected background
and no true signal'', and has been computed
by using the statistical techniques reported in reference~\cite{fc}.
The sensitivity of the proposed experiment has been computed for a
detector of 15~kilotons and a baseline of 735~km.  The resulting sensitivity
plot is shown in figure~\ref{fig-exclu1}.  The systematic uncertainty 
on the \nue\ contamination of the beam is assumed to be of the order of 5\%.

\begin{figure}[ht]
      \begin{center}\mbox{\epsfxsize=14cm\epsffile{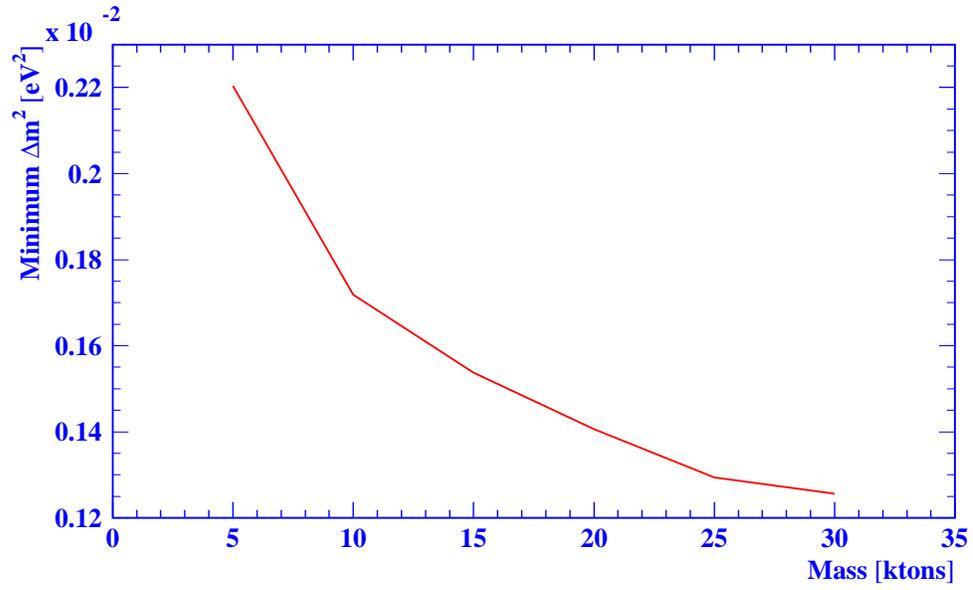}}
      \end{center}
      \caption{Minimum \dm\ at 90\%~C.L. as a function of the mass
               for 735~km}
      \label{fig-mass}
\end{figure}

We studied the possible improvements that can be achieved by varying
the mass and the baseline.  Figure~\ref{fig-mass} shows the lowest \dm\ 
reachable by the experiment  at 735~km as a function of its mass.  
Since the experiment is not background-free, the minimum \dm\ value
does not scale as $1/\sqrt{\rm mass}$.

\begin{figure}[ht]
      \begin{center}\mbox{\epsfxsize=14cm\epsffile{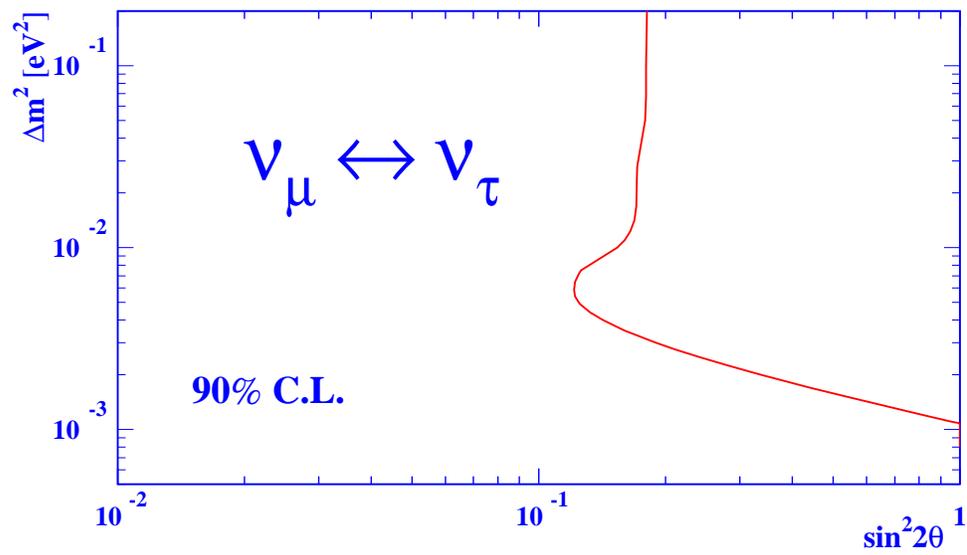}}
      \end{center}
      \caption{Sensitivity plot at 3000~km}
      \label{fig-exclu2}
\end{figure}

When $\dm({\rm eV}^2) \ll E({\rm GeV})/L({\rm km})$, the oscillation
probability $\cal{P}$ is such that the gain in the number of oscillated 
\nutau\ compensates exactly for the loss of neutrino flux $\phi(\numu)$
as the baseline is increased:
${\cal P}(\numtonut)\times\phi(\numu) \simeq {\rm constant}$.
On the other hand, the background is proportional to the $\nu$ flux,
which decreases like $1/L^2$.  
The minimum \dm\ value that can be
probed with a given experiment having a non-negligible background
thus decreases by increasing the distance from the neutrino source,
at the price of a reduced maximum sensitivity.  Figure~\ref{fig-exclu2}
shows the sensitivity plot for the same detector in the same beam,
when it is positioned 3000~km away from the neutrino source.  It 
should be stressed that the systematic uncertainties on the background
(dominated by \nue\ interactions) are in this case less significant 
with respect to the 735~km baseline experiment, since the signal to 
background ratio improves by a factor~$(3000/735)^2\sim 17$.

\section{Conclusion}
We believe that the liquid scintillator detector technology could be
used to build a large detector to search for \numtonut\ oscillations 
in the atmospheric region with a long baseline beam.  Whenever 
$\dm({\rm eV}^2) \ll E({\rm GeV})/L({\rm km})$, with the current
beam designs, the quasi-elastic regime has a significant contribution
to the overall interaction rate of the oscillated \nutau.  In this 
study, we have concentrated on a simple selection of \nutau\
quasi-elastic interactions where the $\tau$ subsequently decays to 
an electron. 
A 15~kiloton detector running for four years at 735~km 
from the neutrino source of a beam similar to the proposed NGS, can probably 
achieve a minimum \dm\ of about 1.5$\times 10^{-3}$~eV$^2$ at 90\% C.L. in 
appearance mode.  This result could improve down to about 
1.1$\times 10^{-3}$~eV$^2$ if the baseline were increased to 3000~km. 
Additional channels in the quasi-elastic regime, as well as the analysis
of deep inelastic events, could improve the potential of the experiment
in both appearance and disappearance modes.

\section*{Acknowledgments}
We would like to thank our CHORUS and I216 colleagues for useful
discussions.  We are grateful to J.-P.~Fabre, P.~Migliozzi and S.~Ricciardi 
for constructive comments and valuable information.  We would also like 
to thank V.~Palladino and N.~Vassilopoulos for providing us with the beam 
spectrum.


\newpage

\end{document}